\documentstyle[aps,prl,multicol]{revtex}
\input{epsf}
\begin{document}
\draft
\preprint{\today}
\title{Level Statistics and Localization 
for Two Interacting Particles in a Random Potential}
\author{Dietmar Weinmann and Jean-Louis Pichard}
\address{CEA, Service de Physique de l'Etat Condens\'e,
         Centre d'Etudes de Saclay,\\
         91191 Gif-sur-Yvette Cedex, France}
\maketitle
\begin{abstract}
 We consider two particles with a local interaction $U$ in 
a random potential at a scale $L_1$ (the one particle localization 
length). A simplified description is provided by a Gaussian 
matrix ensemble with a preferential basis. We define the symmetry 
breaking parameter $\mu \propto U^{-2}$ associated to the 
statistical invariance under change of basis. We show that 
the Wigner-Dyson rigidity of the energy levels is 
maintained up to an energy $E_{\mu}$. We find that $E_{\mu} 
\propto 1/\sqrt{\mu}$ when $\Gamma$ 
(the inverse lifetime of the states of the preferential basis) 
is smaller than $\Delta_2$ (the level spacing), and 
$ E_{\mu} \propto 1/\mu $ when $\Gamma > \Delta_2$. 
This implies that the two-particle localization length 
$L_2$ first increases as $|U|$ before eventually behaving as $U^2$. 
\end{abstract}
\begin{multicols}{2} 
\narrowtext
%
%
 For a single particle diffusing in a disordered system of size $L$ smaller 
than the one particle localization length $L_1$, there are two characteristic 
energies: the Thouless energy $E_{\rm c}=\hbar D/ L^2$ and the level spacing 
$\Delta_1 \approx B_1/ L^d$ ($B_1, D$ and $d$ are the band width, the 
diffusion constant and the system dimension, respectively). If one writes the 
distribution of energy levels as a Gibbs factor of a fictituous Coulomb gas, 
the corresponding pairwise interaction for levels with separation 
$\epsilon < E_{\rm c}$ coincides~\cite{jalpicbeen} with the logarithmic
repulsion characteristic of the matrix ensembles which are statistically 
invariant under change of basis, e. g. the Gaussian Orthogonal Ensemble (GOE). 
For $\epsilon > E_{\rm c}$, the level repulsion vanishes more or less quickly, 
depending on the system dimension. The dimensionless conductance $g_1$ is
given by $E_{\rm c}/\Delta_1$. This ratio is the single relevant parameter 
in the scaling theory of localization. In quasi-one dimension, the size where 
$g_1 \approx 1$ defines $L_1$. In three dimensions, the mobility edge is 
characterized by $g_1 \approx g_{\rm c}$ where $g_{\rm c}$ is of order 1. 

 We shall generalize those concepts to two particles with a local (repulsive or
attractive) interaction. This two interacting particle (TIP) problem has 
received a particular attention since Shepelyansky~\cite{shepelyansky} pointed 
out that certain TIP states may extend over a scale $L_2$ much larger than 
$L_1$. Shepelyansky's original reasoning consists in mapping the problem for 
$L \gg L_1$ onto a random band matrix model with a superimposed 
diagonal matrix (SBRM-model). Imry~\cite{imry} used later the Thouless 
scaling block picture to arrive at precisely the same results as 
Shepelyansky. The smearing due to the
interaction of the energy levels within $L_1$ was estimated using Fermi's 
golden rule, yielding $L_2 \propto U^2$.  This delocalization effect has been 
confirmed by transfer matrix studies~\cite{fmgpw,oppen}, and unambiguously
illustrated from numerical studies~\cite{wmgpf} of rings threaded by an 
$AB$-flux. However, in one dimension, for system sizes which can be 
numerically investigated, one obtains~\cite{oppen} $L_2 \propto |U|$ contrary
to Fermi's golden rule, and a disorder dependence~\cite{fmgpw} 
$L_2 \propto L_1^{\alpha}$ with $\alpha \approx 1.5$ --- $1.7$ and not $2$, 
as predicted by Shepelyansky and Imry. 

  To understand those contradictory results, we study the TIP energy level 
statistics at a scale $L_1$ in order to identify the energy which plays the 
role of $E_{\rm c}$ in this case, and to determine its dependence on $U$. For 
the original TIP-problem, we assume a tight-binding model~\cite{shepelyansky} 
on a $d$-dimensional lattice ($L_1^d$ sites $p$ where the random potential is 
taken with a box distribution of width $2W$). The nearest neighbor hopping 
term takes a constant value $V=1$ and $U$ is the on-site interaction. 
Assuming two electrons with opposite spins, we consider only the 
symmetric states. The TIP-Hamiltonian~\cite{shepelyansky} can be written 
in the basis of the $N=L_1^d(L_1^d+1)/2$ (symmetrized) products of one 
particle states $|AB\rangle $. We denote by $R_{pA}$ the value on site 
$p$ of the one particle eigenstate with energy $\epsilon_{A}$. In this 
basis, the diagonal terms are dominated by one particle contributions 
$\epsilon_{A}+\epsilon_{B}$ and the interaction Hamiltonian yields a 
full matrix (for $L\le L_1$) with entries $U\cdot Q_{ABA'B'}=U 
\sum_p R_{pA}R_{pB}R_{pA'}R_{pB'}$. The magnitude of those terms is 
of order $U/L_1^{3d/2}$ with a random sign which does not preserve the 
sign of the interaction. 

 Before considering the TIP-Hamiltonian, it is instructive to discuss a 
simplified matrix model where the correlations between matrix elements are 
neglected: an ensemble of real symmetric matrices $G$ with independent 
entries, characterized by Gaussian distributions with variances 
$<G_{ii}^2> \approx B_1^2/3 $ ($B_1 = 4Vd+2W$) and 
$<G_{ij}^2> \approx U^2/L_1^{3d}$ for the diagonal and off-diagonal terms, 
respectively. The averages are set to zero, which neglects a shift of the 
diagonal terms by an amount $U\cdot Q_{ABAB} \approx U/L_1^d$ assumed to be 
much smaller than $B_1$. These shifts preserve the sign of $U$ and for large
$U$, eventually split the energy band into two parts. For the sake 
of simplicity, we ignore them, restricting us to small $U$ and to 
a Gaussian matrix with preferential basis (GMPB)-model which has 
been used previously~\cite{haake} to study the GOE to Poisson crossover for 
the level statistics, and to define a maximum entropy 
model~\cite{pichardshapiro} where the range of the level interaction depends 
on a parameter. When $<G_{ii}^2 >$ is very large 
\begin{figure}[tbh]
\epsfxsize=3in
\epsfysize=2.25in
\epsffile{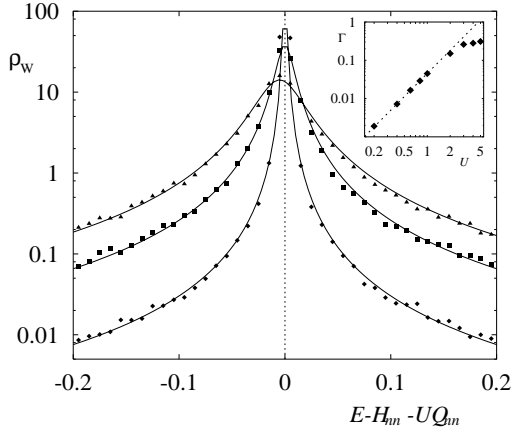}
\vspace{3mm}
\caption[Breit-Wigner]{\label{Breit-Wigner}
The strength function for a TIP-Hamiltonian (5 by 5 lattice in the 
metallic regime $W=2,\quad V=1$). Diamonds, squares and triangles are for 
$U$= 0.2, 0.6 and 1.0, respectively. Lines are Breit--Wigner 
distribution functions, fitted to the numerical data. The inset 
shows how $\Gamma $ depends on $U$. The line represents 
$\Gamma = U^2 /22$.}
\end{figure}
compared to $<G_{ij}^2>$, one
has indeed a strongly preferential basis and it is convenient to re-order the
diagonal terms such that $G_{11} < G_{22}< \ldots <G_{NN}$. Those $G_{ii}$ may 
be considered as the positions of the energy levels in the zeroth 
approximation, when the interaction with the other states is neglected. 
The small coupling terms $G_{ij}$ spread those basis states over $\Gamma/  
\Delta_2$ neighbors. $\Delta_2\approx \sqrt{2 \pi <G_{ii}^2>}/N$ is the level 
spacing and $\Gamma$ can be estimated using Fermi's golden rule: 
$\Gamma \approx 2 \pi <G_{ij}^2> /\Delta_2$.
Diagonalizing $G$ by an orthogonal transformation $O$ 
($G=O G_d O^{t}$, where $G_d$ is a diagonal matrix with real entries 
$E_{\alpha}$), we consider the strength function introduced by Wigner 
$\rho_{\rm W}(E,p)=\sum_{\alpha=1}^N O_{p\alpha}^2 \delta(E-E_{\alpha})$. 
If one averages over the ensemble, one gets for the eigenvector 
amplitudes $<O_{p\alpha}^2>=\Delta_2^{-1} <\rho_{\rm W}(E_{\alpha},p)>$. For 
$L \gg L_1$, $\rho_{\rm W}(E,n)$ has been found~\cite{jacquod,fmg,fyodorov} 
in agreement with the Breit-Wigner form 
\begin{equation}
<\rho_{\rm W}(E,n)>= {\Gamma \over {2\pi [(E-G_{nn})^2+\Gamma^2/4]}}\, .
\end{equation}
We have checked that this holds too for $L \le L_1$, and we show in 
Fig.~1 that this Breit-Wigner form of the eigenstates characterizes 
also the original TIP-Hamiltonian, once the shifts $U\cdot Q_{nn}$ of the 
quasi-energies $H_{nn}$ are taken into account. This shows us that a 
basis state $|n\rangle \equiv |AB\rangle $ (i. e. an eigenstate of the 
TIP-Hamiltonian for $U=0$) becomes delocalized over 
$\Gamma / \Delta_2$ of its neighbors (i.e. over the basis states 
$|n'\rangle $ where $H_{n'n'}$ is close to 
$H_{nn}=\epsilon_{A}+\epsilon_{B}$), with a Lorentzian shape centered in 
$H_{nn}+U\cdot Q_{nn}$. $\Gamma$ plays the role of a localization length in 
the preferential basis, and is given by Fermi's golden rule for small values 
of $U$ ($U \le 2$ in Fig.~1).

 Having understood how the eigenstates are delocalized by the interaction 
over the preferential basis (see also Refs.~\cite{jacquod,flambaum}), 
we focus our attention on the energy levels. We introduce a symmetry 
breaking parameter $\mu$ in the probability density
\begin{equation}\label{density}
\rho(G) \propto \exp\left( -\sum_{i=1}^N {G_{ii}^2 \over {2 \sigma^2}} 
-(1+\mu) \sum_{i<j}^N { G_{ij}^2 \over \sigma^2} \right) \, ,
\end{equation} 
with $\sigma^2\approx B_1^2/3$ and $\sigma^2 /(2(1+\mu))\approx U^2/L_1^{3d}$. 
When $\mu=0$, one recovers the GOE-ensemble with $\rho_{\rm GOE}(G)\propto 
\exp(-tr (G^2)/2\sigma^2)$. When $\mu \neq 0$, there is a factor 
$\rho_{\mu}(G)$ which removes the statistical invariance under change 
of basis. Expressed~\cite{pichardshapiro} in eigenvalue-eigenvector 
coordinates, it reads
\begin{equation}\label{densitymu}
\rho_{\mu} (G) \propto \prod_{\alpha<\beta}^N \exp \left( 
-{\mu \over 2 \sigma^2}  (E_{\alpha}-E_{\beta})^2 
\sum_p O_{p\alpha}^2 O_{p\beta}^2 \right) \, .
\end{equation} 
The question is to understand how this additional factor, after integration 
over the matrices $O$ (distributed with Haar measure $\mu(dO)$ over the 
orthogonal group) can destroy the logarithmic level repulsion coming 
from the measure $\mu(dG)=\prod_{\alpha<\beta}^N | E_{\alpha} - E_{\beta} |
 \prod_{\alpha}^N dE_{\alpha} \mu(dO)$. This will allow us to identify the 
characteristic scale $E_{\mu}$ below which one recovers the GOE rigidity, 
and above which the levels become uncorrelated. 
Two cases have to be considered:

(i) $\Gamma < \Delta_2$. The $G_{ij}$ are so small that one can just 
consider the coupling between two nearest neigbor diagonal entries: i. e. 
a $2\times 2$ matrix which can be diagonalized by a rotation of an angle 
$\theta$. One finds~\cite{pichardshapiro} $\int d\theta 
\rho_{\mu}(G)=f(x)=\exp(-x)\cdot I_0(x)$ where 
$x= \mu\cdot \epsilon^2 / (8 \sigma^2)$, $\epsilon$ denoting the separation 
of the two coupled levels. For $ x < 1$, $f(x)\approx 1$ and decreases as 
$1/\sqrt{x}$ for $x>1$. This gives 
\begin{equation}\label{case1}
{E_{\mu} \over \Delta_2} = {\sqrt{ 8 \sigma^2 /\mu} \over \Delta_2} \propto 
{N \over \sqrt{\mu}}\, .
\end{equation}
For $\epsilon < E_{\mu}$, one has the GOE statistics, while for 
$\epsilon > E_{\mu}$, the levels are uncorrelated.

(ii) $\Gamma > \Delta_2$. Many neighboring $G_{ii}$ are coupled by the 
off-diagonal terms. First, we consider the case where 
$\epsilon=|E_{\alpha}-E_{\beta}| < \Gamma$: i. e. the case where the two 
corresponding eigenvectors have a strong overlap. Assuming that 
the eigenvectors $|O_{\alpha}\rangle$ have non-zero coordinates of order 
$O_{n\alpha}^2 \approx \Delta_2 / \Gamma$ over $\Gamma / \Delta_2$ 
neighboring basis states only, one gets 
$\sum_{p=1}^N O_{p\alpha}^2 O_{p\beta}^2 \approx \Delta_2 / \Gamma$, and 
$\exp \left( -\mu \epsilon^2 \Delta_2/ (2 \sigma^2 \Gamma)\right) 
\approx 1$, independent on $\epsilon (<\Gamma)$. Writing $O=\exp A$, 
with $A$ a real antisymmetric matrix ($\mu(dO)=\prod_{\alpha<\beta} 
dA_{\alpha\beta}$), one can see that the small fluctuations of the 
$A_{n\alpha}$ around their typical values will not yield a correction 
to the GOE level repulsion. This means that there is no coupling 
between eigenvalues and eigenvectors as far as 
$\epsilon < \Gamma \equiv E_{\mu}$ with 
\begin{equation}\label{case2}
{E_{\mu} \over \Delta_2} \propto {N^2 \over \mu}
\end{equation}
\begin{figure}[tbh]
\epsfxsize=3in
\epsfysize=2.25in
\epsffile{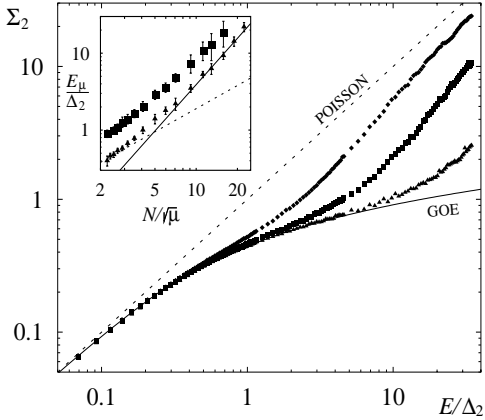}
\vspace{3mm}
\caption[Gaussian matrices]{\label{Gauss}
$\Sigma_2 $ for the GMPB-ensemble ($N=500$). Diamonds, squares and 
triangles are for $\mu$= 30000, 5000 and 1000, respectively. 
The inset shows how $E_{\mu}$ depends on $\mu $. The triangles 
give the energy where $\Sigma_{2}$ is  20 percent 
above the GOE value. The solid and the dotted line represent 
$E_{\mu}/\Delta_2 =0.039 N^2/\mu$ and $E_{\mu}/\Delta_2 =0.19 N/\sqrt{\mu}$, 
respectively. The squares  result from a fit
$\Sigma_{2}=(E/E_{\mu})^{\alpha}$ valid for $E\gg E_{\mu}$.}
\end{figure}
now, instead of $N/\sqrt{\mu}$ previously. When $|\epsilon| > \Gamma$, 
the eigenvectors do not overlap and the levels should become 
uncorrelated. In Ref.~\cite{pichardshapiro}, it was noted that if $O_{p\alpha} 
\approx \delta_{p,\alpha} + A_{p\alpha}$ where $A_{p\alpha}\ll 1$, 
$\sum_{p=1}^N O_{p\alpha}^2 O_{p\beta}^2 \approx 2 A_{\alpha\beta}^2$, 
which gives a $1/|\epsilon|$ factor, after integration over $A_{\alpha\beta}$. 
This level attraction exactly compensates the level repulsion due to 
$\mu(dG)$. Qualitatively, one can adapt this reasoning to produce the 
requested level attraction, after integration over the eigenvectors.
Quantitatively, the calculation of the exact form of the level interaction 
as a function of $\epsilon$, taking into account the Breit-Wigner form of 
the eigenvectors, is postponed to a further study. 

  We have carried out a numerical study of the GMPB-ensemble ($N=500$) 
as a function of $\mu$, to illustrate the two regimes. The number 
variance $\Sigma_2(E)$ (variance of the number of levels  
in an energy interval $E$) is shown in Fig.~2. For small energy 
intervals, $\Sigma_2(E)$ coincides with the GOE-logarithmic increase 
observed when $\mu=0$. For larger energy intervals, $\Sigma_2(E)$ 
can be fitted by $(E/E_{\mu}) ^{\alpha(\mu)}$, which gives a first 
method for calculating $E_{\mu}$. A second method consists in calculating  
the energy interval where $\Sigma_2$ is above the GOE-curve by a certain 
threshold (e. g. 20 percent). Note that those methods give a non zero 
$E_{\mu}$ (depending on the chosen threshold) even for uncorrelated 
levels, which has been subtracted from the data. The inset of Fig.~2 
confirms that the two methods are in agreement and exhibits the 
predicted crossover for $E_{\mu}$ when $\Gamma \approx \Delta_2$ 
($\mu \approx N^2$), from a $N^2/\mu$-dependence (small $\mu$) 
towards a $N/\sqrt{\mu}$-dependence. The $\mu$-dependence of the 
exponent $\alpha$ (see inset in Fig.~4) depends on the exact form of 
the pairwise level repulsion.

  We now turn our attention to the TIP-Hamiltonian in two dimensions. The 
corresponding $\Sigma_2$ is shown in Fig.~3, for different $U$. We have 
obtained the same curves 
\begin{figure}[tbh]
\epsfxsize=3in
\epsfysize=2.25in
\epsffile{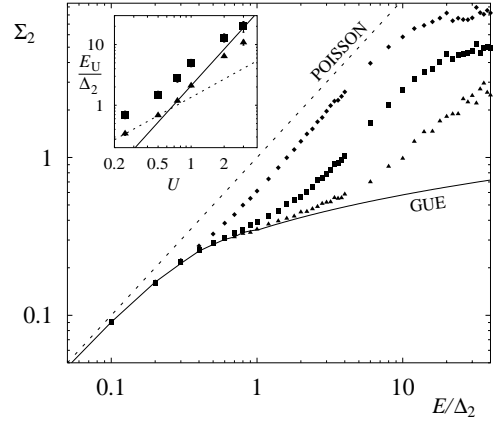}
\vspace{3mm}
\caption[TIP 2D L=10x10]{\label{TIP_2D}
$\Sigma_2$ for the symmetric states of a $2d$ TIP-Hamil\-ton\-ian 
(ring with 10x10 sites, $W=V=1$). The ring is threaded by a magnetic flux 
$\Phi=\Phi_{0}/4$. Diamonds, squares and 
triangles are for $U$= 0.25, 0.75 and 2.0, respectively. 
The inset shows how $E_{\rm U}$ depends on $U$. The data are 
obtained as described in the caption of Fig.~\ref{Gauss}.    
The dotted line (solid line) corresponds to $E_{\rm U}/\Delta_2=1.35 U/V$ 
($E_{\rm U}/\Delta_2=2.1 (U/V)^2$).}
\end{figure}
for $U=1$ and $U=-1$, and consider in more detail 
repulsive interactions. We took rings containing $10 \times 10$ sites 
threaded by a magnetic flux, so that the level statistics should have a GUE 
behavior for energy intervals $E < E_{\rm U} \equiv E_{\mu}$, with 
$ (1+\mu)^{-1} \approx 6 U^2/ (L_1^{3d} B_1^2)$. Except this trivial 
change from orthogonal to unitary symmetry, the similarity with the 
curves calculated for the GMPB-model is very striking, if one disregards 
large energy intervals and the $U$-dependence of the exponent $\alpha$ 
(see inset of Fig.~4). The first difference can be explained from the fact 
that in the TIP Hamiltonian, the one particle level rigidity cannot be 
ignored when $E > \Delta_1$ (one has $\Delta_1 / \Delta_2$ superimposed 
GUE series when $U=0$), correlations which are neglected in the GMPB-model. 
 Another difference results from the 
shifts $U\cdot Q_{nn}$ of the diagonal terms 
which become important when $U$ is 
large. The crossover value $U_{\rm c} \approx {\sqrt 2} 
(8Vd+4W)/(\sqrt{\pi} L^{d/2})$ between the two regimes is of order 1, 
for the considered parameters. When $U>U_{\rm c}$, we observe the 
$U^2$-behavior of $E_{\rm U}$.

 A similar study in one dimension is very instructive. When $W=V=1$, 
we have $L_1\approx 25$, which gives again $U_{\rm c}\approx 1$. As expected, 
one can see in Fig.~4 that $E_{\rm U} \propto |U|$ when $|U| < U_{\rm c}$, but 
when $U > U_{\rm c}$, the splitting of the energy band occurs, and 
$E_{\rm U}$ decreases. Note that one recovers a Poisson statistics 
when $U$ is very large (for $d=2$, there is only a saturation of 
$E_{\rm U}$). For $d=1$, this means that  one can couple only 
two basis states within $L_1 \approx 25$, with a 
small enough value of $U$ to justify the simplified GMPB ensemble. 
The observation of the $U^2$ behavior of $E_{\rm U}$ requires larger 
values of $L_1$ in $d=1$ than 
considered in the numerical studies~\cite{fmgpw,oppen,wmgpf}.     
\begin{figure}
\epsfxsize=3in
\epsfysize=2.25in
\epsffile{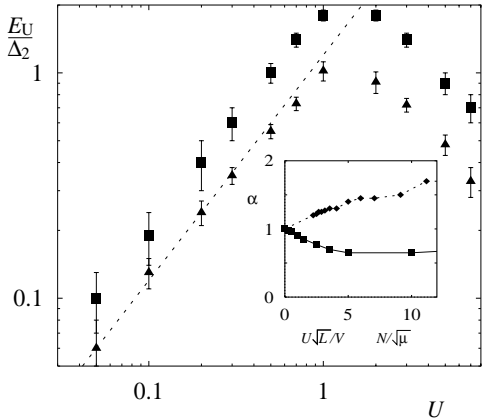}
\vspace{3mm}
\caption[TIP 1D L=25]{\label{TIP_1D}
$E_{\rm U}$ for a $1d$ TIP-Hamiltonian ($W=V=1$ and $L=L_{1}=25$ sites). The 
data are obtained as described in the caption of \ref{Gauss}. The dotted line 
corresponds to $E_{\rm U}/\Delta_2=1.2 U/V$. The inset gives the exponent 
$\alpha$ occurring in $\Sigma_{2}\propto E^{\alpha} $ at large $E$. Squares 
are for the $1d$ TIP Hamiltonian, plotted as a function of $U\sqrt{L}/V$. 
Diamonds represent $\alpha$ for the GMPB ensemble (see Fig.~\ref{Gauss}), as 
a function of $N/\sqrt{\mu}$.} 
\end{figure}

 We now follow the argument developed by Imry~\cite{imry} to estimate 
the localization length $L_2$. First, we consider a series of building 
blocks of size $L_1^d$. $\Gamma_{\rm U}$ is the smearing of the TIP  
levels of one of the blocks, due to the interaction-induced coupling 
with the neighboring block. For such a quasi-$1d$ wire, 
the dimensionless conductance at scale $L_1$ is given by 
\begin{equation}\label{g2}
g_2(L_1)\equiv {L_2 \over L_1} \approx {1 \over 2} + 
A {\Gamma_{\rm U} \over \Delta_2}.
\end{equation} 
The factor $1/2$ gives~\cite{oppen} the right limit when $ U \rightarrow 0$ 
and $A$ is a constant. Obviously, one should have 
$\Gamma_{\rm U}\equiv E_{\rm U}$. When $\Gamma_{\rm U} > \Delta_2$, 
$\Gamma_{\rm U}$ is given by Fermi's golden rule, the case considered in
Ref.~\cite{imry} and we only discuss the case $\Gamma_{\rm U} < \Delta_2$, 
where $\Gamma_U\approx \sqrt{U^2/L_1^{3d}}$ . Physically, this means that $U$ 
is so small that it couples only a single TIP state in one of the blocks to 
another TIP state in the next block, giving rise to ``Rabi oscillations'' 
between those two coupled states. The inverse life time is no longer given 
by the square of the coupling term, as in Fermi's golden rule, but by its 
absolute value. In addition, we have shown that this inverse life time 
gives the scale below which one has a GOE spectral rigidity, extending 
the known results for non-interacting particles to the TIP problem. 

  We conclude by discussing a few implications for TIP localization. For 
$d=1$ and $U<U_{\rm c}$ ($U_{\rm c} \approx 1$ when $W\approx V \approx 
1$, see Fig.~4), one gets $(L_2/ L_1) \approx 1/2 + A ( |U|/ B_1) 
\sqrt{L_1} $ which is in agreement with the dependence on $U$ observed in 
the numerical studies~\cite{oppen}. The conjecture proposed in 
Ref.~\cite{oppen} gives $L_1$ instead of $\sqrt{L_1}$. As noted in 
Ref.~\cite{fmgpw}, the distribution of the $Q_{ABA'B'}$ is far from 
being Gaussian, which can matter as far as the description of the 
$L_1$-dependence by the GMPB-model is concerned. However $ L_2 \propto 
L_{1}^{3/2}$ is close to the behavior observed in Ref~\cite{fmgpw}. 
The $U$--dependence is not affected by this consideration and is 
correctly described by the GMPB-model.

 If one considers two quasi-particles above a Fermi sea, one should 
replace~\cite{imry} in Eq.~(\ref{g2}) $\Delta_2$ by $\Delta_2(E)
\approx \Delta_1^2/E$ where $E$ is the total excitation energy. One 
immediately obtains that the quasi-particle conductance $g^q_2(E,L_1)$ 
is of order of $g_2(L_1)$ when $E\approx B_1$, which gives 
$L^q_2(E\approx B_1)=L_2$, in agreement with Ref.~\cite{oppen2}. 
Similarly, in three dimensions, Imry's relation 
($E_{m2}\approx (B_1^2/|U|) E_{m1}^{\nu d/2}$) between the one  
quasi-particle mobility edge $E_{m1}$ and the two quasi-particle 
mobility edge $E_{m2}$ does not change when $U<U_c$ ($\nu$ denotes 
the critical exponent associated with $L_1$). 

 In summary, we have shown that the basic concepts developed for 
non interacting particles can be naturally extended to $M=2$ 
interacting particles, after the changes $E_{\rm C} \rightarrow E_{\mu}$ 
and $\Delta_1 \rightarrow \Delta_2$. A similar conclusion has been 
obtained from a non linear $\sigma$ model description of the 
TIP-Hamiltonian~\cite{fmgp}, when $L > L_1$. Moreover, our approach  
can be easily extended to an arbitrary number $M$ of particles. 

 We gratefully acknowledge a useful discussion with Boris Shapiro and Dima 
Shepelyansky. This work was supported by the European HCM program (D. W.). 

%

\end{multicols}
\end{document}